\documentclass[twocolumn]{revtex4}
\usepackage[utf8]{inputenc}
\usepackage{graphicx}

\begin{document}

\title{Spin waves and instabilities in the collinear four component antiferromagnetic materials}

\author{Pavel A. Andreev}
\email{andreevpa@my.msu.ru}
\affiliation{Department of General Physics, Faculty of physics, Lomonosov Moscow State University, Moscow, Russian Federation, 119991.}

\date{\today}

\begin{abstract}
The small amplitude perturbations of spins are considered in the four component antiferromagnetic materials
with the equilibrium state of form up-up-down-down (uniaxial samples).
Other configurations for the four component antiferromagnetic materials and two component antiferromagnetic materials are briefly considered for the comparison with main regime.
Dispersion dependencies of two spin waves existing in the system are found if equilibrium spins are parallel to the anisotropy axis.
Dispersion equation leading to a possibility of four spin waves is derived if equilibrium spins are perpendicular to the anisotropy axis.
It is found that at least one solution has negative frequency square for all possible modules and signs of the anisotropy constants.
Calculations are made for the one dimensional chain of classical spins in the approximation of the nearest neighbours interaction.
Next, we also addressed the nearest neighbours interaction approximation in the limit of the continuous medium
(for the Landau--Lifshitz--Gilbert equation).
Mostly applied form of the Landau--Lifshitz--Gilbert equation goes beyond the nearest neighbours interaction approximation.
The difference is described.
Required assumptions are described.
\end{abstract}

%\pacs{}% PACS, the Physics and Astronomy
                             % Classification Scheme.
%\keywords{Heisenberg Hamiltonian}

\maketitle

%%%%%%%%%%TEXT

%\mbox{\boldmath $\sigma$}

\section{Introduction}

Spin waves in the ferromagnetic materials (FM) or the antiferromagnetic materials (AFM) can be considered using
the microscopic quantum Heisenberg equation,
the microscopic quasi-classic Bloch equations for single atoms with the interaction between atoms,
or the macroscopic Landau--Lifshitz--Gilbert equation for continuous medium
\cite{ZvezdinMukhin JETP L 09}, \cite{Gareeva PRB 13}, \cite{Andreev 2025 Vestn}, \cite{Andreev 2025 05}.

Waves can be considered for the collinear equilibrium state \cite{ZvezdinMukhin JETP L 09}, \cite{Gareeva PRB 13},
including the contribution of the Dzylaoshinskii-Moriya interaction \cite{Moon PRB 13}, \cite{Zakeri PRL 10}.
Small amplitude perturbations of the helicoidal \cite{Fishman PRB 19}
or the cycloidal \cite{Fishman PRB 19}, \cite{Andreev 2025 10}, \cite{Andreev 2025 09} spin order can be found as well.

Specific feature of the response of the multiferroic materials
\cite{Katsura PRL 05}, \cite{Sergienko PRB 06}, \cite{Mostovoy npj 24}, \cite{Dong AinP 15}
(here we mean the magnetically ordered materials demonstrating the ferroelectricity of the spin origin)
is the electromagnon resonance.
It reveals in the permittivity multiferroic as a wide resonance (or two close resonances)
in the low frequency are in compare with the antiferromagnetic resonance \cite{Pimenov NP 06}, \cite{ShuvaevPimenov EPJB 11}.
Recent theoretical models for the electromagnon resonance can be found in Refs. \cite{Castro PRB 25}, \cite{AndreevTrukh EPL 25}.

Influence of the Dzylaoshinskii-Moriya interaction \cite{Wang PRL 15}
or the magnetoelectric coupling \cite{Risinggard SR 16} on the nonlinear spin structures is also considered in literature.
While the Dzylaoshinskii-Moriya interaction itself is responsible for the formation of specific nonlinear spin structures such as
skyrmions which possesses additional property called topological charge \cite{Rybakov PRB 19}.

Spin configurations up-up-down-down
can be found in the four-component antiferromagnetic multiferroic materials \cite{Tokura RPP 14}.
Some symmetric properties of the four-component AFM are recently discussed in Ref. \cite{ZvezdinGareeva PSS 24}.
In this paper we are interested in magnetic properties of such materials only
and consider the spin waves in this configuration.
The microscopic quasi-classic Bloch equations is used in our research in the nearest neighbors interaction approximation.
Hence, we obtain the dispersion dependence in the full the Brillouin zone
in contrast with the macroscopic Landau--Lifshitz--Gilbert equation,
which allows to obtain the long-wavelength limit.

The problem of formulation of the exchange interaction in the macroscopic Landau--Lifshitz--Gilbert equation for the AFM
is addressed in literature several times.
An earlier suggestion can be found in review article \cite{Akhiezer UFN 1960}, published in 1960.
Later, it is reexamined to obtain a version mostly used up to date
\cite{Landau 8}, this results were published in 1970th.
Recent derivation of this model with the description of required relation of parameters of the system can be found in Ref. \cite{Andreev 2025 Vestn}.
Both of these models appear to be different from the nearest neighbours interaction approximation.
So, we present the macroscopic Landau--Lifshitz equations and corresponding energy densities
for two-component AFM and for the four-component AFM
obtained in the nearest neighbours interaction approximation.
The next-nearest neighbor exchange parameters are discussed in Ref. \cite{Holbein PRB 23} (see TABLE I) for TbMnO$_{3}$
(as an example useful for magnetoelectric effect study).
The systematic account of next-nearest neighbor interaction with magnitudes different from the nearest neighbor interaction
can be made using intermediate equations of Ref. Ref. \cite{Andreev 2025 Vestn},
but here we specify the regime of complete neglecting next-nearest neighbor interaction.

% \cite{Akhiezer UFN 1960}, )

%In 1974 on Fisher 2306.00955 ref 38    %%  40 and 42
%LL vol 8 eq 49.3 and vol 9 eq 74.8

%electromagnon
%dielectric permeability
%The Dzylaoshinskii-Moriya
%The Heisenberg exchange % magneto-electric coupling,
%the electric dipole moments related to the noncollinear order of spins \cite{Tokura RPP 14}, \cite{Khomskii JETP 21}
%$\textbf{d}_{ij}= \alpha_{ij}[\textbf{r}_{ij}\times[\textbf{s}_{i}\times\textbf{s}_{j}]]$,
%\cite{Sparavigna PRB 94}, \cite{Mostovoy PRL 06}
%$\textbf{P}= \sigma[\textbf{S}(\nabla\cdot \textbf{S})-(\textbf{S}\cdot\nabla)\textbf{S}]$,
%the symmetric Heisenberg Hamiltonian.
%the Landau--Lifshitz--Gilbert equation

This paper is organized as follows.
In Sec. II problem of stability and the dispersion dependence for the four-component AFM is addressed.
In Sec. III the Landau--Lifshitz equation is considered for the four-component AFM
in the nearest neighbours interaction approximation.
In Sec. IV the Landau--Lifshitz equation is considered for the two-component AFM
in the nearest neighbours interaction approximation.
In Sec. V to sum up all described about
we consider possible forms
of the energy density in the nearest-neighbor interaction approximation for the two-component AFM and the four-component AFM.
In Sec. VI
a brief summary of obtained results is presented.

\section{Spin waves in one-dimensional spin chains}

\emph{Model}:

We consider quasiclassic XYZ model for the uniaxial magnetics:
\begin{equation}\label{MFMemf Hamiltonian cl}
\hat{H}\simeq H=-\frac{1}{2}\sum_{i,j,i\neq j}J_{ij}(\textbf{S}_{i}\cdot\textbf{S}_{j})
-\frac{1}{2}\sum_{i,j,i\neq j}(\Delta J_{ij,zz})(S_{i,z}\cdot S_{j,z}).
\end{equation}
In the analysis of dynamical properties of the spin system we restrict ourself with the nearest neighbour interaction regime:
$$\partial_{t}\textbf{S}_{i}(t)=
J_{i,i+1} [\textbf{S}_{i}\times \textbf{S}_{i+1}]
+J_{i,i-1} [\textbf{S}_{i}\times \textbf{S}_{i-1}]$$
\begin{equation}\label{MFMemf spin single evol}
+\tilde{\kappa}_{i,i+1} [\textbf{S}_{i}\times S_{i+1,z}\textbf{e}_{z}]
+\tilde{\kappa}_{i,i-1} [\textbf{S}_{i}\times S_{i-1,z}\textbf{e}_{z}],
\end{equation}
where the shift of one of diagonal values of the exchange integral from the average value
$\Delta J_{ij,zz}$ is represented as the anisotropy coefficient $\tilde{\kappa}_{ij}$
(it is known as the two cite anisotropy).

This quasi-classical model composed of the classical spin vectors of each atom $\textbf{S}_{i}$ is used for the analysis of spin waves.
While the derivation of the macroscopic spin density evolution equation is based on the quantum Hamiltonian,
and presented in the second part of this paper.

\subsection{Spin waves in one-dimensional spin chains -- the easy-axis regime for the four component AFM of form up-up-down-down}

\subsubsection{Regime of two sets of spins of different magnitude}

\begin{figure}\includegraphics[width=8cm,angle=0]{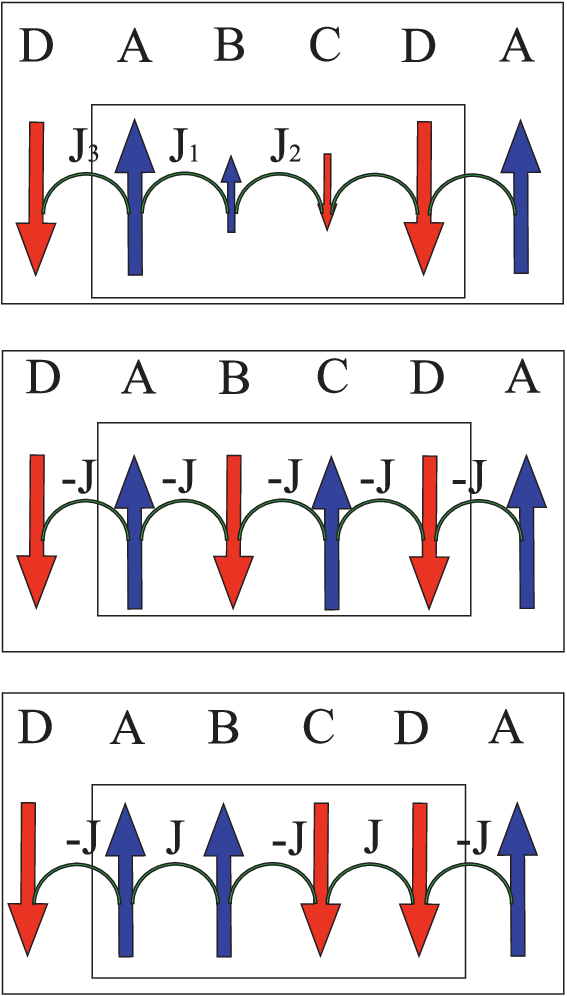}
\caption{\label{MFMannAFM Fig 03}
The equilibrium spin structures considered in this paper and corresponding exchange integrals.
The anisotropy axis is not presented since two regimes are considered:
anisotropy axis parallel to the spins and the anisotropy axis perpendicular to the spins.
} \end{figure}

We consider the collinear equilibrium structure of spins being parallel to the anisotropy axis:
\begin{equation}\label{MFMemf}
  \begin{array}{cccc}
    \textbf{S}_{0,i\in A}=S_{01}\textbf{e}_{z}, & \textbf{S}_{0,i\in B}=S_{02}\textbf{e}_{z},\\
     \textbf{S}_{0,i\in C}=-S_{02}\textbf{e}_{z}, & \textbf{S}_{0,i\in D}=-S_{01}\textbf{e}_{z}. \\
  \end{array}
\end{equation}
This is the configuration of form of up-up-down-down spins.
Corresponding exchange integrals are demonstrated in
Fig. \ref{MFMannAFM Fig 03}.

Perturbations $\textbf{S}_{p}=\textbf{S}_{0,p}+\delta\textbf{S}_{p}$ are considered in the form of plane waves
\begin{equation}\label{MFMemf perturb structure four AFM}
\left(
  \begin{array}{c}
    \delta\textbf{S}_{p} \\
    \delta\textbf{S}_{p+1} \\
    \delta\textbf{S}_{p+2} \\
    \delta\textbf{S}_{p+3} \\
  \end{array}
\right)
=
\left(
  \begin{array}{c}
    \textbf{u}_{1}e^{-\imath\omega t +\imath k p a} \\
    \textbf{v}_{1}e^{-\imath\omega t +\imath k (p+1) a} \\
    \textbf{u}_{2}e^{-\imath\omega t +\imath k (p+2) a} \\
    \textbf{v}_{2}e^{-\imath\omega t +\imath k (p+3) a} \\
  \end{array}
\right),
\end{equation}
where $p$, $p+1$, etc are the numbers of atoms,
$a$ is the distance between neighboring atoms, it is chosen to be equal for all four pairs in the line.

The symmetry of the problem allows to obtain two independent set of equations for $w_{+}=w_{x}+\imath w_{y}$ and $w_{-}=w_{x}-\imath w_{y}$,
where $\textbf{w}$ is one of $\textbf{u}_{1}$, $\textbf{v}_{1}$, etc.
Moreover, the set of equations for $w_{+}$ and $w_{-}$ have same form,
so we can solve one of them to get the full (partially degenerate) dispersion dependence:
\begin{widetext}
\begin{equation}\label{MFMemf}
\left|
\begin{array}{cccc}
  \omega+\Omega_{1}+K_{1} & J_{1}S_{01}e^{\imath ka} & 0 & J_{3}S_{01}e^{-\imath ka} \\
  J_{1}S_{02}e^{-\imath ka} & \omega+\Omega_{2}+K_{2} & J_{2}S_{02}e^{\imath ka} & 0 \\
  0 & -J_{2}S_{02}e^{-\imath ka} & \omega-\Omega_{2}-K_{2} & -J_{1}S_{02}e^{\imath ka} \\
  -J_{3}S_{01}e^{\imath ka} & 0 & -J_{1}S_{01}e^{-\imath ka} & \omega-\Omega_{1}-K_{1}
\end{array}
\right|=0,
\end{equation}
\end{widetext}
where
$\Omega_{1}=J_{3}S_{01}-J_{1}S_{02}$,
$\Omega_{2}=J_{2}S_{02}-J_{1}S_{01}$,
$K_{1}=\tilde{\kappa}_{3}S_{01}-\tilde{\kappa}_{1}S_{02}$,
and
$K_{2}=\tilde{\kappa}_{2}S_{02}-\tilde{\kappa}_{1}S_{01}$.

The dispersion equation can be presented in the following form
$$\omega^{4}-\omega^{2}[\sigma_{1}^{2}+\sigma_{2}^{2}+2J_{1}^{2}S_{01}S_{02}-J_{2}^{2}S_{02}^{2}-J_{3}^{2}S_{01}^{2}]$$
$$+\sigma_{1}^{2}\sigma_{2}^{2}
-2J_{1}^{2}S_{01}S_{02}\sigma_{1}\sigma_{2}
-J_{2}^{2}S_{02}^{2}\sigma_{1}^{2}
-J_{3}^{2}S_{01}^{2}\sigma_{2}^{2}$$
\begin{equation}\label{MFMemf disp dep uudd parall}
+S_{01}^{2}S_{02}^{2}[(J_{1}^{2}-J_{2}J_{3})^{2}+4J_{1}^{2}J_{2}J_{3}\sin^{2}2ka]=0
\end{equation}
where
$\sigma_{i}=\Omega_{i}+K_{i}$.

We see presence of two spin waves (some characteristics are discussed below).
It is in analogy with the two-component AFM,
where single spin wave propagates if the equilibrium spin density is parallel to the anisotropy axis.
So, in both cases there is degeneracy allowing the splitting of the spin wave on two close branches.

\subsubsection{ Simplified regime of four spins of equal magnitude}

To get the main features of the dispersion dependence following from equation (\ref{MFMemf disp dep uudd parall})
we consider some simplifies regimes.
We chose the equal magnitudes of the exchange integrals,
with the signs depending on the relative orientation of neighboring spins.
We chose $J_{i',i'+1}\equiv J>0$ if spins have same direction (in the equilibrium state),
and we chose $J_{i',i'+1}<0$ if spins are antiparallel $J_{i',i'+1}\equiv -J$.
Therefore, we obtain $J_{2}=J_{3}=-J_{1}\equiv -J$,
with $J>0$.
We have similar relation of the relative signs for the anisotropy constants:
$\tilde{\kappa}_{2}=\tilde{\kappa}_{3}=-\tilde{\kappa}_{1}\equiv -\tilde{\kappa}$
with $\tilde{\kappa}> 0$ or $\tilde{\kappa}< 0$,
%{\color{red}(while we can expect $\tilde{\kappa}> 0$ for the easy-axis regime),}
but the subindexless anisotropy constant $\kappa$ can be positive or negative.
These assumptions lead to
$\sigma_{1}=\sigma_{2}\equiv-\sigma_{0}$,
where
$\sigma_{0}=-J(S_{01}+S_{02})-\tilde{\kappa}(S_{01}+S_{02})$.

We can also represent the equilibrium spins in the following form
$S_{01}\equiv S_{0}$,
and $S_{02}\equiv \beta S_{0}$.
Made assumption lead to the simplified form of the dispersion dependence (\ref{MFMemf disp dep uudd parall}):
$$\omega^{2}=\frac{1}{2}\biggl[2\sigma_{0}^{2}-(1-\beta)^{2}J^{2}S_{0}^{2}$$
\begin{equation}\label{MFMemf}
\pm\sqrt{16\beta J^{2}S_{0}^{2}\sigma_{0}^{2} -16\beta^2 J^{4}S_{0}^{4}\sin^{2}2ka+(1-\beta)^{4}J^{4}S_{0}^{4}}\biggr].
\end{equation}

Consider the center of the
Brillouin zone
$\omega(k=0)$,
for $\beta=1$ or $S_{01}=S_{02}$:
\begin{equation}\label{MFMemf}
\omega^{2}(k=0)=\left(
                  \begin{array}{c}
                    4(JS_{0}+K)(2JS_{0}+K) \\
                    2K(JS_{0}+K) \\
                  \end{array}
                \right).
\end{equation}
Assuming that the anisotropy is relatively small $K <JS_{0}$
we have $(JS_{0}+K)>0$
and $(2JS_{0}+K)>0$.
Therefore, we need
$K>0$
to get the stable system.

\begin{figure}\includegraphics[width=8cm,angle=0]{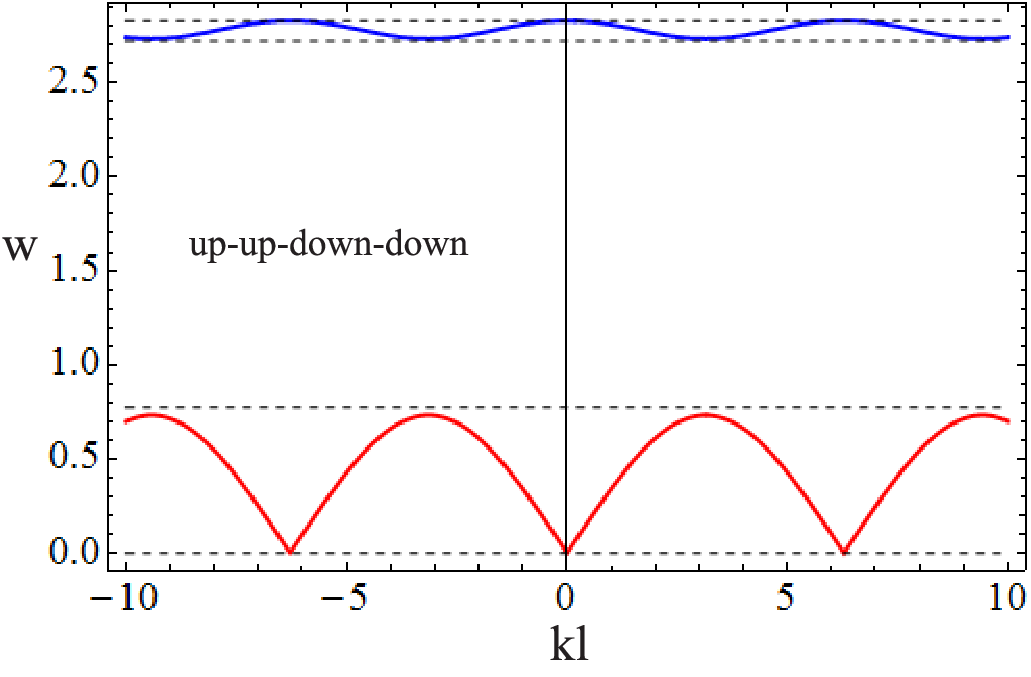}
\caption{\label{MFMannAFM Fig 05} The dispersion dependence of the spin waves existing in four component up-up-down-down system,
if the equilibrium spins are parallel to the anisotropy axis.
It is presented as the function of the dimensionless frequency $\textrm{w}=\omega/JS_{0}$ depending on the dimensionless wave vector $kl$,
with $l=4a$ and $a$ is the interparticle distance.
This figure demonstrates functions given by equation (\ref{MFMemf dispersion dependence uudd no an}).} \end{figure}

%\textbf{ww}
In the limit of equal modulus of spins $\beta=1$
we get the dimensionless form of the dispersion dependence for two spin waves
(at the additional assumption of the zero anisotropy energy):
\begin{equation}\label{MFMemf dispersion dependence uudd no an}
\frac{\omega}{JS_{0}}=2\sqrt{1\pm\sqrt{1-\frac{1}{4}\sin^{2}(kl/2)}},
\end{equation}
with $l=4a$.
Dispersion dependencies (\ref{MFMemf dispersion dependence uudd no an}) are demonstrated in Fig. \ref{MFMannAFM Fig 05}.

We also present the behavior of the dispersion dependencies near the center of the Brillouin zone
\begin{equation}\label{MFMemf}
\omega=JS_{0}\left(
                \begin{array}{c}
                  \sqrt{2}ka, \\
                  2\sqrt{2}(1-\frac{1}{8}(ka)^2), \\
                \end{array}
              \right)
\end{equation}
where we see characteristic frequency for the upper branch $2\sqrt{2}JS_{0}$.

The lowest branch has the linear dispersion dependence,
which can be additionally shifted up due to the anisotropy energy contribution.
The upper branch has the quadratic dispersion dependence with the negative group velocity
(from further comparison we can conclude that the group velocity is relatively small),
with the additional shift up due to the anisotropy energy contribution.

\subsection{Regime of easy-axis for the four component AFM with configuration up-down-up-down}

\begin{figure}\includegraphics[width=8cm,angle=0]{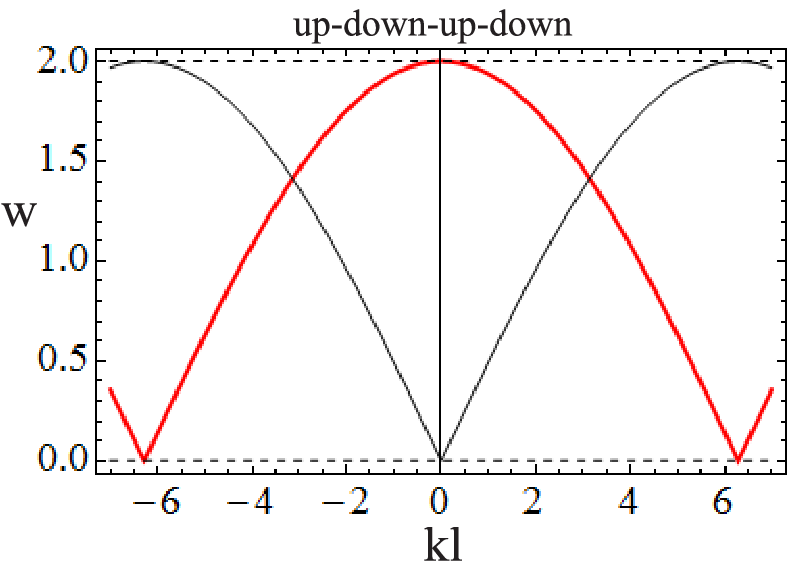}
\caption{\label{MFMannAFM Fig 05b} The dispersion dependence of the spin waves existing in four component up-down-up-down system,
if the equilibrium spins are parallel to the anisotropy axis.
The dimensionless notations are the same as in previous figure.
This figure illustrated equation (\ref{MFMemf disp dep 4 AFM udud ea}).} \end{figure}

We need something to compare with the results found in previous subsection.
So, instead of the four component AFM with configuration up-up-down-down
we consider the four component AFM with configuration up-down-up-down
(without the anisotropy energy).
We choose equal magnitudes of spins.
We also assume that all exchange integrals negative and equal to each other.
The small amplitude perturbations of form of equation (\ref{MFMemf perturb structure four AFM}).
It leads to the following dispersion dependence
\begin{equation}\label{MFMemf}
\biggl(\frac{\omega^{2}}{J^{2}S_{0}^{2}}-2\biggr)^{2}-4\cos^{2}2ka=0.
\end{equation}
Corresponding dispersion dependence is presented in Fig. \ref{MFMannAFM Fig 05b}.
Its analytical form is
\begin{equation}\label{MFMemf disp dep 4 AFM udud ea}
\frac{\omega}{JS_{0}} =2 \left(
                  \begin{array}{c}
                    \cos kl/4, \\
                    \sin kl/4, \\
                  \end{array}
                \right)
\end{equation}
with $l=4a$.

This regime (up-down-up-down) shows the lower value of the frequency of the upper branch of the dispersion curve at the center
of the Brillouin zone:
$\omega(k=0)\mid_{upper}=2JS_{0}$,
in comparison with the previous regime,
where we get $2\sqrt{2}JS_{0}$.
Moreover, in this regime (up-down-up-down) we see that the frequency of both branches changes in interval $\omega\in [0,2JS_{0}]$,
while in the previous subsections each branch changes in the relatively narrow intervals of frequencies.

\begin{figure}\includegraphics[width=8cm,angle=0]{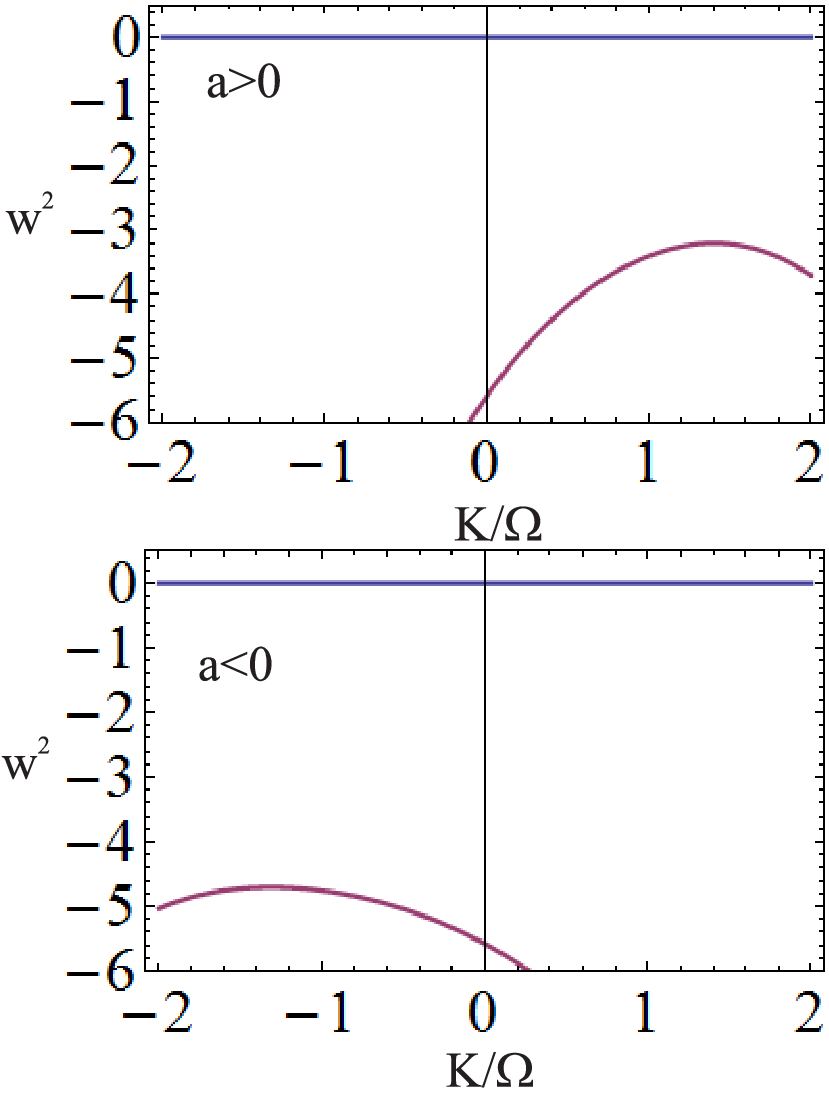}
\caption{\label{MFMannAFM Fig 06}
The frequency square as the function of the dimensionless anisotropy constant
is demonstrated
for the lowest branch of the dispersion dependence following from equation
(\ref{MFMemf disp eq easy plane uudd}).
} \end{figure}

\subsection{Spin waves in one-dimensional spin chains -- the easy plane regime for the two component AFM}

In order to get some expectations for the easy-plane regime in the four-component AFM
we consider the two-component AFM.
In the easy-axis regime
(macroscopic interaction constant $\kappa=\kappa_{AA}=-\kappa_{AB}=>0$,
see discussion after equation (\ref{MFMemf LLG for S A in four comp}),
see also Ref. \cite{Andreev 2025 10},
and $\textbf{S}_{0}\parallel \textbf{e}_{z}$,
where $\textbf{e}_{z}$ is the direction of the anisotropy axis)
we get one spin wave
(its dispersion dependence is described in the recent Ref. \cite{Andreev 2025 10}
in the long-wavelength limit).
In the easy-plane regime ($\kappa<0$ and $\textbf{S}_{0}\perp \textbf{e}_{z}$)
we get two stable spin waves with the following dispersion properties
(their long-wavelength limit is also described in Ref. \cite{Andreev 2025 10}).

In the easy-plane regime, with the equilibrium of form of
$\left(
  \begin{array}{cc}
    \textbf{S}_{0,i\in A}=S_{0}\textbf{e}_{x}, & \textbf{S}_{0,i\in B}=-S_{0}\textbf{e}_{x}     \\
  \end{array}
\right)$,
we have no symmetry between $w_{+}$ and $w_{-}$.
So, we can deal with $\textbf{u}_{1}$, $\textbf{v}_{1}$.
It leads to the following dispersion dependence
\begin{widetext}
\begin{equation}\label{MFMemf}
\left|
  \begin{array}{cccc}
    \imath\omega & -\tilde{\Omega} & 0 & -(\tilde{\Omega}+\tilde{K})\cos ka \\
    \tilde{\Omega} & \imath\omega & \tilde{\Omega} \cos ka & 0 \\
    0 & (\tilde{\Omega}+\tilde{K})\cos ka & \imath\omega & \tilde{\Omega} \\
    -\tilde{\Omega} \cos ka & 0 & -\tilde{\Omega} & \imath\omega \\
  \end{array}
\right|=0,
\end{equation}
\end{widetext}
where $\tilde{\Omega}=2JS_{0}$ and $\tilde{K}=2\tilde{\kappa} S_{0}$.
It leads to
$$(\omega^{2}-\tilde{\Omega}^{2})^{2}
+2\tilde{\Omega}(\tilde{\Omega}+\tilde{K})\cos^{2}ka(\omega^{2}-\tilde{\Omega}^{2})$$
\begin{equation}\label{MFMemf}
+\tilde{\Omega}^{2}\cos^{2}ka[(\tilde{\Omega}+\tilde{K})^2 \cos^{2}ka -\tilde{K}^2]=0.
\end{equation}

%{\color{red}may be it would be better via $\omega^{2}$ ??}

Consider the center of the
Brillouin zone
$\omega(k=0)$:
\begin{equation}\label{MFMemf}
\omega^{2}(k=0)=\left(
                  \begin{array}{c}
                    0 \\
                    -2\tilde{\Omega}\tilde{K} \\
                  \end{array}
                \right).
\end{equation}
For comparison with the macroscopic model
we have
$\tilde{\Omega}\sim g_{0u}\sim J_{AB}<0$,
see discussion after equation (\ref{MFMemf LLG for S A in four comp}),
see also Ref. \cite{Andreev 2025 10},
and we also have
$\tilde{K}\sim\tilde{\kappa}=\tilde{\kappa}_{AB}\sim\kappa_{AB}>0$ for the anisotropy contribution.

\subsection{Spin waves in one-dimensional spin chains -- the easy-plane regime for the four component AFM with configuration up-up-down-down}

In this regime we chose that the equilibrium spins are parallel to each other,
but the plane of their location is perpendicular to the anisotropy axis:
\begin{equation}\label{MFMemf}
  \begin{array}{cccc}
    \textbf{S}_{0,i\in A}=S_{01}\textbf{e}_{x}, & \textbf{S}_{0,i\in B}=S_{02}\textbf{e}_{x},\\
     \textbf{S}_{0,i\in C}=-S_{02}\textbf{e}_{x}, & \textbf{S}_{0,i\in D}=-S_{01}\textbf{e}_{x}. \\
  \end{array}
\end{equation}
We consider the dynamics of the small amplitude perturbations,
which have form similar to equation (\ref{MFMemf perturb structure four AFM}).

It leads to the dispersion equation,
which can be presented in the following form
\begin{equation}\label{MFMemf disp eq easy plane uudd}
D_{0}+2D_{4}\cos4ka+2D_{8}\cos8ka =0,
\end{equation}
where
three coefficients
depend on the frequency square
and the parameters of the system,
they have the following explicit form
\begin{equation}\label{MFMemf}
D_{8}%=D_{-8}
=\Omega_{1}^{2}\Omega_{2}^{2}K_{1}^{2}K_{2}^{2},
\end{equation}

\begin{widetext}
$$D_{4}%=D_{-4}
=-\omega^{4}(\Omega_{1}^{2}K_{2}^{2}+\Omega_{2}^{2}K_{1}^{2})$$
$$+2\omega^{2}\biggl[\Omega_{0}^{2}\biggl( (\Omega_{1}\Omega_{2}-K_{1}K_{2})(\Omega_{1}K_{2}-\Omega_{2}K_{1})
+2\Omega_{1}\Omega_{2}K_{1}K_{2}\biggr)
+(\Omega_{1}\Omega_{2}-K_{1}K_{2})(\Omega_{2}K_{1}^{3}-\Omega_{1}K_{2}^{3})\biggr]$$
\begin{equation}\label{MFMemf}
-\Omega_{0}^{4}(\Omega_{1}^{2}\Omega_{2}^{2}+K_{1}^{2}K_{2}^{2})
+2\Omega_{0}^{2}\biggl( K_{1}^{2}K_{2}^{2}(\Omega_{1}^{2}+\Omega_{2}^{2}) +\Omega_{1}^{2}\Omega_{2}^{2}(K_{1}^{2}+K_{2}^{2})\biggr)
-\Omega_{1}^{2}\Omega_{2}^{2}(K_{1}^{4}+K_{2}^{4})-K_{1}^{2}K_{2}^{2}(\Omega_{1}^{4}+\Omega_{2}^{4}),
\end{equation}
%\end{widetext}
with
$\Omega_{0}=\Omega_{1}-\Omega_{2}$,
and
%\omega^{2}-\Omega_{0}^{2}
%\begin{widetext}
$$D_{0}=(\omega^{2}-\Omega_{0}^{2})^4
+2(\Omega_{0}^{2}+2(\Omega_{2}K_{2}-\Omega_{1}K_{1}))(\omega^{2}-\Omega_{0}^{2})^3$$
$$+2\biggl( \Omega_{0}^{2}[2(\Omega_{2}K_{2}-\Omega_{1}K_{1})
-(\Omega_{1}^{2}+\Omega_{2}^{2}+K_{1}^{2}+K_{2}^{2})]
+3(\Omega_{1}^{2}K_{1}^{2}+\Omega_{2}^{2}K_{2}^{2})
-4\Omega_{1}\Omega_{2}K_{1}K_{2}\biggr)(\omega^{2}-\Omega_{0}^{2})^2$$
$$+\biggl[ \Omega_{0}^{2}\biggl( -10\Omega_{1}\Omega_{2}K_{1}K_{2} +8(\Omega_{1}^{2}K_{1}^{2}+\Omega_{2}^{2}K_{2}^{2})
-4(\Omega_{1}^{2}K_{2}^{2}+\Omega_{2}^{2}K_{1}^{2})
+4(\Omega_{1}^{3}K_{1}+\Omega_{1}K_{1}^{3}-\Omega_{2}^{3}K_{2}-\Omega_{2}K_{2}^{3})\biggr)$$
$$+2\biggl(\Omega_{2}^{3}K_{2}^{3}-\Omega_{1}^{3}K_{1}^{3}
+\Omega_{1}\Omega_{2}K_{1}K_{2}(\Omega_{1}K_{1}-\Omega_{2}K_{2})\biggr)\biggr](\omega^{2}-\Omega_{0}^{2})$$

$$+\Omega_{0}^{4}\biggl(\Omega_{1}^{4}+\Omega_{2}^{4}+K_{1}^{4}+K_{2}^{4} -2\Omega_{1}\Omega_{2}K_{1}K_{2}
+6(\Omega_{1}^{2}K_{1}^{2}+\Omega_{2}^{2}K_{2}^{2})
+4(\Omega_{1}^{3}K_{1}+\Omega_{1}K_{1}^{3}-\Omega_{2}^{3}K_{2}-\Omega_{2}K_{2}^{3})\biggr)$$

$$+2\Omega_{0}^{2}\biggl( 2\Omega_{1}\Omega_{2}K_{1}K_{2} (\Omega_{1}K_{1}-\Omega_{2}K_{2})
+2 (\Omega_{2}^{3}K_{2}^{3}-\Omega_{1}^{3}K_{1}^{3})
-(\Omega_{1}^{4}+\Omega_{2}^{4})(K_{1}^{2}+K_{2}^{2})-(\Omega_{1}^{2}+\Omega_{2}^{2})(K_{1}^{4}+K_{2}^{4})\biggr)$$
\begin{equation}\label{MFMemf}
+(\Omega_{1}^{4}+\Omega_{2}^{4})(K_{1}^{4}+K_{2}^{4})+2\Omega_{1}^{2}\Omega_{2}^{2}K_{1}^{2}K_{2}^{2}.
\end{equation}
\end{widetext}

Dispersion equation
(\ref{MFMemf disp eq easy plane uudd})
is analyzed in different regimes.
All of them show the negative value of the frequency square for the lowest branch of the dispersion dependence (one of four branches).
It shows the instability of the chosen equilibrium state under the small amplitude perturbations.

It is known that the dispersion dependence of the perturbations of the cycloid equilibrium spin order
(the cycloid is located in the plane which is perpendicular to the anisotropy axis) has form similar to the dispersion dependence of the collinear equilibrium order of spins being in the same plane \cite{Andreev 2025 10}.
The cycloid order presented by the wave vector of the static cycloid modifies the coefficients of the dispersion dependence obtained in the collinear regime \cite{Andreev 2025 10}.

Some systematic modification of the coefficients made in our research does not allow to make the discussed above frequency square positive.
Hence, we expect that the cycloidal order of spin can be unstable as well.
And this instability related to another effect rather then cycloid formation.

For the illustration of the found behavior of the system we plot Fig. \ref{MFMannAFM Fig 06},
where the frequency square is considered for the zero wave vector $\omega^{2}(k=0)$, i.e. the center of the Brillouin zone.
In Fig. \ref{MFMannAFM Fig 06} we assume $\Omega_{2}=-\Omega_{1}$, with $\Omega_{1}>0$ and $\mid K_{2}\mid=\mid K_{1}\mid$, but $K_{2}=aK_{1}$,
with $a=\pm1$, and $K_{1}>0$ and $K_{1}<0$.
Hence we get all possibilities 1) $K_{1}>0$, $K_{2}>0$; 2) $K_{1}>0$, $K_{2}<0$; 3) $K_{1}<0$, $K_{2}>0$; 4) $K_{1}<0$, $K_{2}<0$.
Hence, we obtain the dependence of the frequency square (at $k=0$) as the function of $K_{1}$ for two regimes $a>0$ and $a<0$.
All of them show the strong instability.

%Main consequence is given in Fig. \ref{MFMannAFM Fig 06}

\section{Landau--Lifshitz equation for the four-component AFM}
%via partial spin densities

The problem of stability of the spin configuration is important itself.
However, the analysis presented above reminds us
that the
nearest neighbor interaction
approximation is the basic approach to the study of waves of elements on the lattice or in chain.
One can deliberately consider the interaction with atoms (ions/elements) located further.
This interaction is assumed to be smaller,
so we need to keep it in mind at the construction of the macroscopic models.

The macroscopic models, such as the
Landau--Lifshitz--Gilbert equation,
have their own well-known limitations.
Regarding spin waves,
the macroscopic approach allows to find the long-wavelength limit of the dispersion dependence
(the main term and several corrections to it, if it is necessary).

If we consider the macroscopic model for the contribution of the exchange interaction in AFM existed in period of 1940-1970 (approximately)
\cite{Akhiezer UFN 1960},
we get an expression different from one considered later
\cite{Landau 8}.
More or less, these expressions are based on the symmetry analysis of the macroscopic properties of the medium.
On the other hand, these is well-known approach
allowing to represent the microscopic model (like (\ref{MFMemf Hamiltonian cl}) and (\ref{MFMemf spin single evol}))
in the form of
macroscopic
Landau--Lifshitz--Gilbert
equation (see Ref. \cite{KOSEVICH PR 90}).
Was it used for the systematic examination of the suggested macroscopic models and their consistency with the
nearest neighbor interaction
approach?
Existence of this method is mentioned in Ref. \cite{Akhiezer UFN 1960},
but no clear discussion of its application is given.

Below we use another approximation for the derivation of the macroscopic equations:
the quantum hydrodynamic method.
The quantum hydrodynamic method allows to find macroscopic equations in proper form of the material fields,
but it is less sensitive to the
nearest neighbor interaction approach.
Nevertheless, the
nearest neighbor interaction
approach
can be partially traced in the quantum hydrodynamics for the AFM
since we deal with the interaction of different species.

We consider the four component AFM,
so we can derive equations for the evolution of the partial spin densities corresponding to the spin of each of four subspecies in the system.
To demonstrate main issue we present the spin density of one subspecies.
The derivation is made using the many-particle quantum hydrodynamic method.
It is developed in Ref. \cite{MaksimovTMP 2001}.
Some technical details can be found in Ref. \cite{Andreev 2025 Vestn}.
Additional application of the quantum hydrodynamic method to the magnetic materials can be found
in Refs. \cite{AndreevTrukh JETP 24}, \cite{AndreevTrukh PS 24}, \cite{Andreev 2025 11}, \cite{Andreev 2025 12}.
In the derivation we use the quantum analog of classic Hamiltonian (\ref{MFMemf Hamiltonian cl}).
Possible (collinear) spin configurations are presented in Fig. (\ref{MFMannAFM Fig 03}).
We see that subspecies $A$ is surrounded with subspecies $B$ and $D$,
so they contribute in all parts of the exchange interaction (including the anisotropy energy).
The
Landau--Lifshitz--Gilbert equation
for subspecies $A$ in the
nearest neighbours interaction approach
has the following form
$$\partial_{t}\textbf{S}_{A}
=g_{0,u,AB}  \textbf{S}_{A}\times \textbf{S}_{B}
+g_{0,u,AD}  \textbf{S}_{A}\times \textbf{S}_{D}$$
$$+\kappa_{AB} \textbf{S}_{A}\times S_{B}^{z}\textbf{e}_{z}
+\kappa_{AD} \textbf{S}_{A}\times S_{D}^{z}\textbf{e}_{z}$$
\begin{equation}\label{MFMemf LLG for S A in four comp}
+\frac{1}{6}
g_{2,u,AB} \textbf{S}_{A}\times \triangle \textbf{S}_{B}
+\frac{1}{6}
g_{2,u,AD} \textbf{S}_{A}\times \triangle \textbf{S}_{D}.\end{equation}
Here we deal with the interaction constants
$g_{0,u,AB}\equiv\int J(r)d^{3}r$
$g_{2,u,AB}\equiv\int r^2J(r)d^{3}r$
$\kappa_{AB}\equiv g_{0,\kappa,AB}\equiv \int \tilde{\kappa}(r)d^{3}r$,
which are integral characteristics of the exchange integral and
the shift of one of diagonal values of the exchange integral from the average value
$\Delta J_{ij,zz}\equiv\kappa_{ij}$
presented in equation (\ref{MFMemf Hamiltonian cl}).
Detailed introduction of the interaction constants can be found in Ref. \cite{Andreev 2025 Vestn}.

If we consider AFM it is traditional to consider the full "magnetization" and antiferromagnetic vectors.
For the two-component AFM there is one antiferromagnetic vector,
while for the four-component AFM there are three antiferromagnetic vectors:
\begin{equation}\label{MFMemf}
  \begin{array}{c}
    \textbf{F}\equiv \textbf{M}=\textbf{S}_{1}+\textbf{S}_{2}+\textbf{S}_{3}+\textbf{S}_{4}, \\
    \textbf{A}\equiv \textbf{L}_{1}=\textbf{S}_{1}-\textbf{S}_{2}-\textbf{S}_{3}+\textbf{S}_{4}, \\
    \textbf{G}\equiv \textbf{L}_{2}=\textbf{S}_{1}-\textbf{S}_{2}+\textbf{S}_{3}-\textbf{S}_{4},  \\
    \textbf{C}\equiv \textbf{L}_{3}=\textbf{S}_{1}+\textbf{S}_{2}-\textbf{S}_{3}-\textbf{S}_{4},  \\
  \end{array}
\end{equation}
with
$\textbf{S}_{1}=\textbf{S}_{A}$, $\textbf{S}_{2}=\textbf{S}_{B}$,
$\textbf{S}_{3}=\textbf{S}_{C}$, and
$\textbf{S}_{4}=\textbf{S}_{D}$.

We also present the reverse expressions
\begin{equation}\label{MFMemf}
  \begin{array}{c}
    \textbf{S}_{A}=(\textbf{M}+\textbf{L}_{1}+\textbf{L}_{2}+\textbf{L}_{3})/4, \\
    \textbf{S}_{B}=(\textbf{M}-\textbf{L}_{1}-\textbf{L}_{2}+\textbf{L}_{3})/4, \\
    \textbf{S}_{C}=(\textbf{M}-\textbf{L}_{1}+\textbf{L}_{2}-\textbf{L}_{3})/4,  \\
    \textbf{S}_{D}=(\textbf{M}+\textbf{L}_{1}-\textbf{L}_{2}-\textbf{L}_{3})/4,  \\
  \end{array}
\end{equation}
where we see the same structure except the value of coefficients.

We can represent equation (\ref{MFMemf LLG for S A in four comp})
via $\textbf{M}$, $\textbf{L}_{1}$, $\textbf{L}_{2}$, and $\textbf{L}_{3}$,
but it leads to a rather huge equation if we do not specify the relative values of the interaction constants
$g_{0,u,AB}$ and $g_{0,u,AD}$,
$g_{2,u,AB}$ and $g_{2,u,AD}$,
$\kappa_{AB}$ and $\kappa_{AD}$.
Moreover, we expect to get different relations for different configurations presented in Fig. \ref{MFMannAFM Fig 03}.
We consider some specific configurations for the energy density below.

\section{Landau-Lifshitz equation for two-component AFM}

% Landau--Lifshitz--Gilbert equation

Here we consider some features of the Landau--Lifshitz equation
for the two-component AFM
in the nearest-neighbor interaction approximation.
First, we note that the microscopic derivation
of the Landau-Lifshitz equation
discussed in the works
\cite{Andreev 2025 Vestn}
and \cite{AndreevTrukh PS 24}.
It was assumed that there was an interaction with a second row of neighboring atoms having the same spin projection,
it was assumed to be nonzero, having the same modulus, but the opposite sign of the interaction constant
(a similar concept was considered in \cite{Andreev 2025 11}, \cite{Andreev 2025 12}).
This made it possible to obtain the closest possible form of equations in relation to those presented in
\cite{Landau 8}. %
Features associated with different coupling constants are discussed \cite{Andreev 2025 Vestn}.
Physically, the difference in coupling constants can be due to the distance between atoms (nearest neighbors or second-row neighbors)
or the difference in atoms/ions (electron configuration of ions with different charges of the same nucleus) in a ferrimagnet.

Taking into account the above,
we write down the system
Landau--Lifshitz equations
for a two-component antiferromagnet in the nearest-neighbor interaction approximation
$$\partial_{t}\textbf{S}_{A}
=g_{0,u,AB}  \textbf{S}_{A}\times \textbf{S}_{B}
+\kappa_{AB} \textbf{S}_{A}\times S_{B}^{z}\textbf{e}_{z}$$
\begin{equation}\label{MFMemf}
+\frac{1}{6}
g_{2,u,AB} \textbf{S}_{A}\times \triangle \textbf{S}_{B},\end{equation}
where we assume
$g_{0,u,AA}=0$,
$\kappa_{AA}=0$,
$g_{2,u,AA}=0$,
$\kappa_{2,AA}=0$
in contrast with Ref. \cite{Andreev 2025 Vestn},
which addressed the possibility of rederivation of equations presented in \cite{Landau 8}.
%$g_{2,(\gamma),AA}=0$
%$g_{2,(\beta),AA}=0$

The equation of the spin evolution for the second subspecies is
$$\partial_{t}\textbf{S}_{B}
=g_{0,u,AB}  \textbf{S}_{B}\times \textbf{S}_{A}
+\kappa_{AB} \textbf{S}_{B}\times S_{A}^{z}\textbf{e}_{z}$$
\begin{equation}\label{MFMemf}
+\frac{1}{6}
g_{2,u,AB}\textbf{S}_{B}\times \triangle \textbf{S}_{A},\end{equation}
where the interaction constants with subindexes "BB" is assumed to be equal to zero as well.

Moreover, we present equations
for the antiferromagnetic vector
$\textbf{L}=\textbf{S}_{A}-\textbf{S}_{B}$
and "magnetization" $\textbf{M}=\textbf{S}_{A}+\textbf{S}_{B}$:
$$\partial_{t}\textbf{L}=g_{0,u,AB}\textbf{L}\times\textbf{M}
+\frac{1}{2}\kappa_{AB}(\textbf{L}\times \textbf{e}_{z}M_{z}-\textbf{M}\times\textbf{e}_{z}L_{z})$$
\begin{equation}\label{MFMemf eq L evol 2 comp}
+\frac{1}{12}g_{2,u,AB}(\textbf{L}\times\triangle\textbf{M}-\textbf{M}\times\triangle\textbf{L})
,\end{equation}
and
$$\partial_{t}\textbf{M}=
\frac{1}{2}\kappa_{AB}(-\textbf{L}\times \textbf{e}_{z}L_{z}+\textbf{M}\times\textbf{e}_{z}M_{z})$$
\begin{equation}\label{MFMemf eq M evol 2 comp}
+\frac{1}{12}g_{2,u,AB}(\textbf{M}\times\triangle\textbf{M}-\textbf{L}\times\triangle\textbf{L})
.\end{equation}

The nearest neighbor interaction
approximation presented here leads to the additional terms
in comparison with the "symmetric" case including
the next-nearest neighbor interaction with the atoms/ions of the same species.

Moreover, all terms in equations (\ref{MFMemf eq L evol 2 comp}) and (\ref{MFMemf eq M evol 2 comp}) have different sign
in compare with equations in Ref. \cite{Andreev 2025 Vestn},
where the negative value of the exchange integral between spins of the opposite direction is included explicitly
(other signs, like the anisotropy energy, is "normalized" in the same way).

\section{Energy density in the nearest-neighbor interaction approximation}

We also write down the energy density for a two-component antiferromagnetic materials
in the nearest-neighbor interaction approximation
$$\mathcal{E}_{A}=-\frac{1}{2}g_{0,u,AB}(\textbf{S}_{A}\cdot\textbf{S}_{B})
-\frac{1}{2}\kappa_{0,AB}(S_{A}^{z}\cdot S_{B}^{z})$$
\begin{equation}\label{MFMemf en density XYZ A nni}
-\frac{1}{12}g_{2,u,AB}(\textbf{S}_{A}\cdot\triangle\textbf{S}_{B}).
\end{equation}
Expression for the second subspecies can be found from
(\ref{MFMemf en density XYZ A nni}) as
$$\mathcal{E}_{B}=\mathcal{E}_{A}(A\leftrightarrow B).$$
Next, the full energy density can be found
$$\mathcal{E}=\mathcal{E}_{A}+\mathcal{E}_{B}.$$
As the result, the final expression
for the full energy density is presented via
the antiferromagnetic vector
$\textbf{L}=\textbf{S}_{A}-\textbf{S}_{B}$
and "magnetization" $\textbf{M}=\textbf{S}_{A}+\textbf{S}_{B}$:
$$\mathcal{E}=\frac{1}{4}g_{0,u,AB}(\textbf{L}\cdot\textbf{L}-\textbf{M}\cdot\textbf{M})+\frac{1}{4}\kappa_{0,AB}(L^{z}\cdot L^{z}-M^{z}\cdot M^{z})$$
\begin{equation}\label{MFMemf density XYZ AandB nni beyond}
+\frac{1}{24}g_{2,u,AB}(\textbf{L}\cdot\triangle\textbf{L}-\textbf{M}\cdot\triangle\textbf{M}).
\end{equation}

\subsection{Two component regimes for the energy density: account of next-nearest neighbor interaction}
%Fisher regime and term beyond the nnI

For comparison, we write down the energy density for a two-component antiferromagnet
in the presence of interaction with a second row of neighbors
under the assumption that the module of the interaction constants of ions with the same or different spin projections are equal
\cite{Andreev 2025 12}, \cite{Andreev 2025 Vestn}

$$\mathcal{E}_{A}=-\frac{1}{2}g_{0,u,AA}(\textbf{S}_{A}\cdot\textbf{S}_{A})-\frac{1}{2}g_{0,u,AB}(\textbf{S}_{A}\cdot\textbf{S}_{B})$$
$$-\frac{1}{2}\kappa_{0,AA}(S_{A}^{z}\cdot S_{A}^{z})-\frac{1}{2}\kappa_{0,AB}(S_{A}^{z}\cdot S_{B}^{z})$$
\begin{equation}\label{MFMemf density XYZ A nni beyond}-\frac{1}{12}g_{2,u,AA}(\textbf{S}_{A}\cdot\triangle\textbf{S}_{A})
-\frac{1}{12}g_{2,u,AB}(\textbf{S}_{A}\cdot\triangle\textbf{S}_{B}).
\end{equation}
Expression for the second subspecies $B$ can be found from
(\ref{MFMemf density XYZ A nni beyond}) as
$\mathcal{E}_{B}=\mathcal{E}_{A}(A\leftrightarrow B)$,
hence the full energy density can be found
$\mathcal{E}=\mathcal{E}_{A}+\mathcal{E}_{B}$.

If we apply a common assumption for the relation between the exchange integrals between subspecies
\begin{equation}\label{a}
U_{AB}=-U_{AA}=-U_{BB}\equiv -U
\end{equation}
we get corresponding simplified full energy density
$$\mathcal{E}=-\frac{1}{2}g_{0,u}(\textbf{L}\cdot\textbf{L})-\frac{1}{2}\kappa_{0}(L^{z}\cdot L^{z})$$
\begin{equation}\label{MFMemf density XYZ AandB nni beyond}
-\frac{1}{12}g_{2,u}(\textbf{L}\cdot\triangle\textbf{L}).
\end{equation}
Here, we also redefine some coefficients:
$A\equiv g_{2,u}/6$ for the exchange constant,
and $\kappa\equiv \kappa_{0}$ for the anisotropy constant.

%\subsection{Two component regimes for the energy density: nnI-nearest neighbor interaction}

%nearest neighbor interaction

\subsection{Four component regimes for the energy density: nearest neighbor interaction}

\subsubsection{Configuration up-up-down-down}

Here we consider the energy density for the
four-component AFM with configuration
up-up-down-down of parallel spins.
First, we present the partial energy density $\mathcal{E}_{A}$ in terms of the partial spin densities $\textbf{S}_{A}$, $\textbf{S}_{B}$, etc:
$$\mathcal{E}_{A}=-\frac{1}{2}g_{0,u,AB}\textbf{S}_{A}\cdot\textbf{S}_{B}-\frac{1}{2}g_{0,u,AD}\textbf{S}_{A}\cdot\textbf{S}_{D}$$
$$-\frac{1}{2}\kappa_{0,AB}S_{Az}S_{Bz}-\frac{1}{2}\kappa_{0,AD}S_{Az}S_{Dz}$$
\begin{equation}\label{MFMemf}
+\frac{1}{12}g_{2,u,AB}\partial^{\beta}\textbf{S}_{A}\cdot\partial^{\beta}\textbf{S}_{B}+\frac{1}{12}g_{2,u,AD}\partial^{\beta}\textbf{S}_{A}\cdot\partial^{\beta}\textbf{S}_{D},
\end{equation}
where we make the representation
$-\frac{1}{12}g_{2,u,AB}\textbf{S}_{A}\cdot\triangle\textbf{S}_{B}-\frac{1}{12}g_{2,u,AD}\textbf{S}_{A}\cdot\triangle\textbf{S}_{D}
\rightarrow
\frac{1}{12}g_{2,u,AB}\partial^{\beta}\textbf{S}_{A}\cdot\partial^{\beta}\textbf{S}_{B}+\frac{1}{12}g_{2,u,AD}\partial^{\beta}\textbf{S}_{A}\cdot\partial^{\beta}\textbf{S}_{D}$,
which is possible at the consideration of the full energy.

Relative signs and modules of the exchange integrals (and corresponding, but unnecessary, relations for the anisotropy coefficients)
are demonstrated in Fig. \ref{MFMannAFM Fig 03} (the lower picture).
So, we obtain
$g_{0,u,AB}=g_{0,u}>0$, $g_{0,u,AD}=-g_{0,u}$, $g_{2,u,AB}=g_{2,u}>0$, $g_{2,u,AD}=-g_{2,u}$,
$\kappa_{0,AB}=\kappa_{0}$, and $\kappa_{0,AD}=-\kappa_{0}$.
It leads to the following full energy density:
$$\mathcal{E}_{\Sigma}=\mathcal{E}_{A}+\mathcal{E}_{B}+\mathcal{E}_{C}+\mathcal{E}_{D}$$
$$=
\frac{1}{4}g_{0,u}(\textbf{L}_{1}^{2}-\textbf{L}_{3}^{2})
+\frac{1}{4}\kappa_{0}(L_{1z}L_{1z}-L_{3z}L_{3z})$$
\begin{equation}\label{MFMemf}
+\frac{1}{24}g_{2,u}(\partial^{\beta}\textbf{L}_{1}\cdot\partial^{\beta}\textbf{L}_{1}-\partial^{\beta}\textbf{L}_{3}\cdot\partial^{\beta}\textbf{L}_{3}),
\end{equation}
where we find the contribution of two of four antiferromagnetic vectors.

\subsubsection{Configuration up-down-up-down}

Next, we describe
the energy density for the
four-component AFM with configuration
up-down-up-down of parallel spins.
The partial energy density $\mathcal{E}_{A}$ in terms of the partial spin densities $\textbf{S}_{A}$, $\textbf{S}_{B}$ etc
has structure presented above,
but we have different relation of signs of the exchange integrals  (see Fig. \ref{MFMannAFM Fig 03}, the middle picture),
where we find
$g_{0,u,AB}=g_{0,u,AD}=-g_{0,u}$ with $g_{0,u}>0$, $g_{2,u,AB}=g_{2,u,AD}=-g_{2,u}$ with $g_{2,u}>0$,
and $\kappa_{0,AB}=\kappa_{0,AD}=-\kappa_{0}$.
It leads to the following full energy density:
$$\mathcal{E}_{\Sigma}=\mathcal{E}_{A}+\mathcal{E}_{B}+\mathcal{E}_{C}+\mathcal{E}_{D}$$
$$=
\frac{1}{4}g_{0,u}(\textbf{M}^{2}-\textbf{L}_{2}^{2})
+\frac{1}{4}\kappa_{0}(M_{z}M_{z}-L_{2z}L_{2z})$$
\begin{equation}\label{MFMemf}
+\frac{1}{24}g_{2,u}(\partial^{\beta}\textbf{M}\cdot\partial^{\beta}\textbf{M}-\partial^{\beta}\textbf{L}_{2}\cdot\partial^{\beta}\textbf{L}_{2}),
\end{equation}
where we find the final expression in terms of two of four antiferromagnetic vectors,
but in this regime we obtain the different pair.

\subsection{Discussion}

Definitely, the complete analysis of physical behavior requires
the account next-nearest neighbor interaction
in addition to the nearest neighbor interaction.
Mostly, the contribution of the next-nearest neighbor interaction
is small in comparison to the nearest neighbor interaction.
So, systematic account of the next-nearest neighbors with smaller coefficients in possible for the macroscopic models
such as the Landau--Lifshitz--Gilbert equation.
However, we decided to distinguish the nearest neighbor interaction contribution in the equations presented above in comparison with the well-known macroscopic approach presented in the well-known textbooks.

%nearest neighbor interaction

\section{Conclusion}

Small amplitude perturbations and their dispersion dependence is the fundamental characteristics of the system.
It allow to
complete
the analysis of the equilibrium state problem.
It gives the check of the stability of the possible equilibrium.
Instability of the up-up-down-down parallel configuration of spins if they are oriented perpendicular to the anisotropy axis in the uniaxial samples has been found.
Dispersion dependencies for the up-up-down-down and up-down-up-down parallel configurations of spins are found and compared.
Some distinctive differences have been pointed out.

Moreover,
the Landau--Lifshitz--Gilbert equation is the fundamental macroscopic tool for the description of the collective spin dynamics.
The assumptions underlying this equation have been examined for the AFM.
The performed principal examination is based on the microscopic justification of the Landau--Lifshitz equation,
and their relation to the nearest neighbor interaction approximation.

\section{DATA AVAILABILITY}

%\emph{DATA AVAILABILITY}:
Data sharing is not applicable to this article as no new data were
created or analyzed in this study, which is a purely theoretical one.

\section{Acknowledgements}

%\emph{Acknowledgements}:
The work is supported by the Russian Science Foundation under the
grant
No. 25-22-00064.

%\appendix

%\section{Set for LLG}{\color{red}for B}  {\color{red} for C for D }

%\newpage

%\section{Supplementary materials}


\begin{thebibliography}{17}



\bibitem{ZvezdinMukhin JETP L 09} A. K. Zvezdin, A. A. Mukhin,
"On the effect of inhomogeneous magnetoelectric (flexomagnetoelectric) interaction on the
spectrum and properties of magnons in multiferroics",
JETP Lett. \textbf{89}, \textbf{328}, 332 (2009).
%http://link.springer.com/10.1134/S0021364009070042


\bibitem{Gareeva PRB 13}
Z. V. Gareeva, A. F. Popkov, S. V. Soloviov, and A. K. Zvezdin,
"Field-induced phase transitions and phase diagrams in BiFeO$_{3}$-like multiferroics"
Phys. Rev. B \textbf{87}, 214413 (2013).
%DOI: 10.1103/PhysRevB.87.214413


\bibitem{Andreev 2025 Vestn} P. A. Andreev,
"Generalization and microscopic justification of the material-field form of the Landau-Lifshitz equation for antiferromagnets",
Moscow University Physics Bulletin \textbf{80}, 959 (2025).
%DOI: 10.3103/S0027134925703229



\bibitem{Andreev 2025 05} P. A. Andreev,
"Regimes of optical transparency and instabilities of collinear dielectric ferromagnetic materials in the presence of the dynamic magnetoelectric effect",
Phys. Scr. \textbf{100}, 125962 (2025).
%arXiv:2505.07107



\bibitem{Moon PRB 13} J.-H. Moon, S.-M. Seo, K.-J. Lee, K.-W. Kim, J. Ryu, H.-W. Lee, R. D. McMichael, and M. D. Stiles,
"Spin-wave propagation in the presence of interfacial Dzyaloshinskii-Moriya interaction" ,
Phys. Rev. B \textbf{88}, 184404 (2013).

\bibitem{Zakeri PRL 10} K. Zakeri, Y. Zhang, J. Prokop, T.-H. Chuang, N. Sakr, W. X. Tang, and J. Kirschner,
"Asymmetric Spin-Wave Dispersion on Fe(110):
Direct Evidence of the Dzyaloshinskii-Moriya Interaction",
Phys. Rev. Lett. \textbf{104}, 137203 (2010).


\bibitem{Fishman PRB 19}
R. S. Fishman, T. R\~{o}\~{o}m, and R. de Sousa,
"Normal modes of a spin cycloid or helix",
Phys. Rev. B \textbf{99}, 064414 (2019).
%DOI: 10.1103/PhysRevB.99.064414



\bibitem{Andreev 2025 10} P. A. Andreev,
"Electric susceptibility of antiferromagnetic multiferroics with cycloidal spin order at magnetoelectric effect associated with collinear component of spins",
JETP Letters \textbf{123}, 124 (2026).
%JETP Letters, 2026, Vol. 123, No. 2, pp. 124
%DOI: 10.1134/S0021364025610322
%arXiv:2510.07425.


\bibitem{Andreev 2025 09} P. A. Andreev,
"Analytical analysis of the spin wave dispersion in the cycloidal spin structures under
the influence of magneto-electric coupling",
arXiv:2509.19543
(accepted to Phys. Scr.)
%10.1088/1402-4896/ae5696


\bibitem{Katsura PRL 05} H. Katsura, N. Nagaosa, and A. V. Balatsky,
"Spin Current and Magnetoelectric Effect in Noncollinear Magnets",
Phys. Rev. Lett. \textbf{95}, 057205 (2005).
%DOI: 10.1103/PhysRevLett.95.057205

\bibitem{Sergienko PRB 06} I. A. Sergienko, E. Dagotto,
"Role of the Dzyaloshinskii-Moriya interaction in multiferroic perovskites", Phys. Rev. B \textbf{73}, 094434 (2006).


\bibitem{Mostovoy npj 24} M. Mostovoy,
"Multiferroics: different routes to magnetoelectric coupling",
npj Spintronics, \textbf{2} 18 (2024).
%https://doi.org/10.1038/s44306-024-00021-8

\bibitem{Dong AinP 15}
S. Dong, J.-M. Liu, S.-W. Cheong, Z. Ren,
"Multiferroic materials and magnetoelectric physics: symmetry, entanglement, excitation, and topology",
Advances in Physics \textbf{64}, 519 (2015).
%DOI: 10.1080/00018732.2015.1114338
%64:5-6, 519-626,



\bibitem{Pimenov NP 06}
A. Pimenov, A. Mukhin, V. Ivanov, V. D. Travkin, A. M. Balbashov, and A. Loidl,
"Possible evidence for electromagnons in multiferroic manganites",
Nature Phys. \textbf{2}, 97 (2006).
%p. 97вЂ“100
%https://doi.org/10.1038/nphys212



%\bibitem{Pimenov JP CM 08} A. Pimenov, A. M. Shuvaev, A. A. Mukhin and A. Loidl,
%"Electromagnons in multiferroic manganites", J. Phys.: Condens. Matter \textbf{20}, 434209 (2008).
%DOI 10.1088/0953-8984/20/43/434209



\bibitem{ShuvaevPimenov EPJB 11} A. M. Shuvaev, F. Mayr, A. Loidl, A. A. Mukhin, and A. Pimenov,
"High-frequency electromagnon in GdMnO$_{3}$",
Eur. Phys. J. B \textbf{80}, 351 (2011).
%Eur. Phys. J. B 80, 351вЂ“354 (2011)
%1008.2064v1
%https://doi.org/10.1140/epjb/e2011-10691-3



\bibitem{Castro PRB 25} M. A. Castro, C. Saji, G. Saez, P. Vergara, S. Allende, and A. S. Nunez,
"Phenomenological theory of electromagnons in multiferroic systems", Phys. Rev. B \textbf{111}, 214401 (2025).
%DOI: https://doi.org/10.1103/PhysRevB.111.214401
%Mario A. Castro, Carlos Saji, Guidobeth Saez, Patricio Vergara, Sebastian Allende, and Alvaro S. Nunez







\bibitem{AndreevTrukh EPL 25} P. A. Andreev, M. I. Trukhanova,
"Mean-field theory of the electromagnon resonance",
EPL \textbf{152}, 56001 (2025).


\bibitem{Wang PRL 15}
W. Wang, M. Albert, M. Beg, M.-A. Bisotti, D. Chernyshenko,
D. Cortes-Ortuno,
I. Hawke, H. Fangohr,
"Magnon-Driven Domain-Wall Motion with the Dzyaloshinskii-Moriya Interaction" ,
Phys. Rev. Lett. \textbf{114}, 087203 (2015).
%DOI: 10.1103/PhysRevLett.114.087203




\bibitem{Risinggard SR 16} V. Risinggard, I. Kulagina, J. Linder,
"Electric field control of magnoninduced magnetization dynamics in multiferroics",
Scientific Reports \textbf{6}, 31800 (2016).
%DOI: 10.1038/srep31800 1


\bibitem{Rybakov PRB 19}
F. N. Rybakov, and N. S. Kiselev
"Chiral magnetic skyrmions with arbitrary topological charge",
Physical Review B \textbf{99}, 064437 (2019).
%DOI: 10.1103/PhysRevB.99.064437


\bibitem{Tokura RPP 14} Y. Tokura, S. Seki, and N. Nagaosa,
"Multiferroics of spin origin",
Rep. Prog. Phys. \textbf{77}, 076501 (2014).
%(45pp)
%doi:10.1088/0034-4885/77/7/076501



\bibitem{ZvezdinGareeva PSS 24} A. K. Zvezdin, Z. V. Gareeva,
"Symmetry analysis of conductive antiferromagnetic materials CuMnAs, Mn$_{2}$Au",
Physics of the Solid State, 2024, Vol. 66, No. 6, P. 784
%DOI: 10.61011/PSS.2024.06.58685.42HH

%{\color{red}\bibitem{Fisher PRL 74}on Fisher 2306.00955 ref 38, 40 and 42}


\bibitem{Akhiezer UFN 1960}
A. I. Akhiezer, V. G. Bar'yakhtar, M. I. Kaganov,
"Spin waves in ferromagnets and antiferromagnets. I",
Physics–Uspekhi \textbf{3}, 567 (1961).
%Physics–Uspekhi, 1961, Volume 3, Issue 4, Pages 567–592
%DOI: https://doi.org/10.1070/PU1961v003n04ABEH003309
%ufn12363



\bibitem{Landau 8}
L. D. Landau, E. M. Lifshitz,
Second Edition (1984).
\emph{Electrodynamics of Continuous Media}.
Volume 8 in Course of Theoretical Physics. ergamon Press Ltd., Headington Hill Hall, Oxford OX3 0BW, England.
%https://doi.org/10.1016/B978-0-08-030275-1.50003-5

\bibitem{Holbein PRB 23}
S. Holbein, P. Steffens, S. Biesenkamp, J. Ollivier, A. C. Komarek, M. Baum, and M. Braden,
"Spin-wave dispersion and magnon chirality in multiferroic TbMnO$_{3}$",
Phys. Rev. B \textbf{108}, 104404 (2023).
%10.1103/PhysRevB.108.104404
%10.48550/arXiv.2308.09407


\bibitem{KOSEVICH PR 90}
A. M. Kosevich, B. A. Ivanoy and A. S. Kovalev,
"Magnetic Solitons",
Physics Reports %(Review Section of Physics Letters)
\textbf{194}, 117 (1990).

\bibitem{MaksimovTMP 2001} L. S. Kuz'menkov, S. G. Maksimov, and V. V. Fedoseev,
"Microscopic quantum hydrodynamics of systems of
fermions: Part I," Theoretical and Mathematical
Physics \textbf{126}, 110 (2001).


\bibitem{AndreevTrukh JETP 24} P. A. Andreev, M. I. Trukhanova,
"Equation of evolution of electric polarization of multiferroics proportional to the vector
product of spins of ions of the cell under the influence of the Heisienberg Hamiltonian",
JETP, \textbf{166}, 665 (2024) [in russian].
%665вЂ“678,


\bibitem{AndreevTrukh PS 24} P. A. Andreev, M. I. Trukhanova,
"Electric polarization evolution equation for antiferromagnetic
multiferroics with the polarization proportional to the scalar product of the spins",
Phys. Scr. \textbf{99}, 1059b2 (2024).

\bibitem{Andreev 2025 11} P. A. Andreev,
"On the generalized Keffer form of the Dzyaloshinskii constant:
its consequences for the spin, momentum and polarization evolution",
arXiv:2511.21672.

\bibitem{Andreev 2025 12} P. A. Andreev,
"Keffer-like form of the symmetric Heisenberg exchange integral:
Contribution to the Landau--Lifshitz--Gilbert equation and spin wave dispersion dependence",
arXiv:2512.22108.













%\bibitem{Sergienko PRL 06} I. A. Sergienko, C. Sen, and E. Dagotto,
%"Ferroelectricity in the Magnetic E-Phase of Orthorhombic Perovskites", Phys. Rev. Lett. \textbf{97}, 227204 (2006).
%10.1103/PhysRevLett.97.227204





%\bibitem{Khomskii JETP 21} D. I. Khomskii,
%"Multiferroics and Beyond: Electric Properties of Different Magnetic Textures", JETP \textbf{132}, 482 (2021).
%Journal of Experimental and Theoretical Physics


%\bibitem{Sparavigna PRB 94} A. Sparavigna, A. Strigazzi, and A. Zvezdin,
%"Electric-field effects on the spin-density wave in magnetic ferroelectrics", Phys. Rev. B \textbf{50}, 2953 (1994).
%10.1103/PhysRevB.50.2953


%\bibitem{Mostovoy PRL 06} M. Mostovoy, "Ferroelectricity in Spiral Magnets",
%Phys. Rev. Lett. \textbf{96}, 067601 (2006).
%DOI: 10.1103/PhysRevLett.96.067601



%\bibitem{Solovyev PRL 21} I. Solovyev, R. Ono, and S. Nikolaev,
%"Magnetically induced polarization in centrosymmetric bonds", Phys. Rev. Lett. \textbf{127}, 187601 (2021).
%DOI:https://doi.org/10.1103/PhysRevLett.127.187601




%\bibitem{AndreevTrukh EPJ B 24} P. A. Andreev, M. I. Trukhanova,
%"Polarization evolution equation for exchange-strictionally formed type II multiferroic materials",
%Eur. Phys. J. B \textbf{97}, 116 (2024).
%https://doi.org/10.1140/epjb/s10051-024-00756-7
%"Polarization evolution equation for the exchange-strictionally formed II-type multiferroic materials",
%{\color{red}arXiv:2312.16321.}


%\bibitem{JuraschekBalatsky PRM 17} D. M. Juraschek, M. Fechner, A. V. Balatsky, and N. A. Spaldin,
%"Dynamical multiferroicity", Phys. Rev. Mater. \textbf{1}, 014401 (2017).




%\bibitem{Mahajan PRL 11} S. M. Mahajan, F. A. Asenjo, Phys. Rev. Lett. \textbf{107}, 195003 (2011).





%-----------------------------------------------------------------

%%%%%%%%%%%% to SUPPLEMENTARY MATERIALS only


%Nos. 3 and 4, 117вЂ”238 (1990).



\end{thebibliography}
\end{document}